\documentclass[aps,prl,twocolumn,groupedaddress,showpacs]{revtex4}
\usepackage{graphicx}
\usepackage{dcolumn}
\usepackage{bm}
\usepackage{color}%
\usepackage{amsmath}
\usepackage{amsfonts}

\begin{document}

\title{Adsorption properties of a nitrogen atom on the anionic golden fullerene Au$_{16}^{-}$}

\author{Gunn Kim}
\email[e-mail: ]{gunnkim@sejong.ac.kr}
\affiliation{Department of Physics, Graphene Research Institute and Institute of Fundamental Physics, Sejong University, Seoul 143-747, Korea}

\author{Chan-young Lim}
\affiliation{Sejong Science High School, Seoul 152-881, Korea}

\author{Seoung-Hun Kang}
\author{Young-Kyun Kwon}
\email[e-mail: ]{ykkwon@khu.ac.kr}
\affiliation{Department of Physics and Research Institute for Basic Sciences, Kyung Hee University, Seoul 130-701, Korea}

\date{\today}

\begin{abstract}
Using density functional theory, we examine a nitrogen-doped
anionic golden cage (NAu$_{16}^-$).
For the exohedral adsorption that is more stable than the endohedral doping,
the bridge and hollow sites have larger binding energies than the atop sites by $\sim$1 eV.
When the N atom is adsorbed on the cage, electrons are transferred from Au$_{16}^-$ to the N atom.
The transition between the exohedral and endohedral adsorption
may occur thermally through a bridge site. In the infrared active
vibrational spectra, exohedral doping  causes greater intensities at higher
frequencies than endohedral doping.
\end{abstract}


\pacs{61.48.−c, 68.43.Bc, 73.22.-f}

\maketitle

\section{Introduction}
\label{Introduction}

Since their discovery, carbon fullerene molecules such as C$_{60}$,
C$_{70}$, and C$_{80}$ have attracted considerable attention because of
their remarkable electronic, optical, and electrochemical
properties~\cite{HIVFD,FFFD,EDFD,ESFD}. Interestingly, it was proved
experimentally that carbon fullerenes have cousins made of metal, in
particular, golden fullerenes. In 2006, S. Bulusu and colleagues showed
experimental and theoretical evidence of anionic hollow golden cages,
Au$_{n}^-$ ($n=16-18$)~\cite{Bulusu06}. They compared experimental
photoelectron spectra of the anionic structures with those simulated by
density functional theory (DFT)~\cite{Kohn}.

Au$_{16}^-$ is the smallest golden cage with an empty interior. It has
slightly distorted tetrahedral symmetry ($T_d$), with an inner diameter
of ${\approx}5.5$~\AA. Thus, it is possible to endohedrally dope the
golden cage in analogy to the carbon fullerenes. Numerous theoretical
and experimental studies of golden cages adsorbed with metal atoms have
been reported~\cite{PCCP06,JCP07Si,ANGE07CU,PLA08}. Similar to carbon
fullerene derivatives, this kind of golden cage can be a host to form
various derivatives by adsorbing various atoms, molecules and
radicals. Recently, due to their high reactivity, gold nanoclusters have
been explored for their potential use for nanocatalysts~\cite{Yoon03,
Valden98,Heiz00,Boyen02}.

In this paper, we report our study on a golden cage anion, Au$_{16}^-$,
doped with a nitrogen atom which is a non-metal atom~\cite{SHKang11}.
Since oxidation is an important issue for gold nanostructures because of
their high reactivity, we investigate the binding characteristics of
the adsorbate atom on Au$_{16}^-$. Nitrogen is an element with a large
electron affinity, and so the adsorbed N atom may accept electrons from
the golden fullerene. Going a step further, it may help to understand
adsorption properties of molecules containing nitrogen such as N$_2$,
NO, NO$_2$, and NH$_3$, onto the golden buckyball. We also investigate
the transition between the most favorable exohedral and endohedral
adsorption configurations. Finally, we show the infrared (IR) active
vibrational spectra of our models, which may help identify N-adsorbed
Au$_{16}^-$ in future experiments.

\section{Computational details}
\label{Computational}

\begin{figure*}[t]
\centering
\includegraphics[width=1.8\columnwidth]{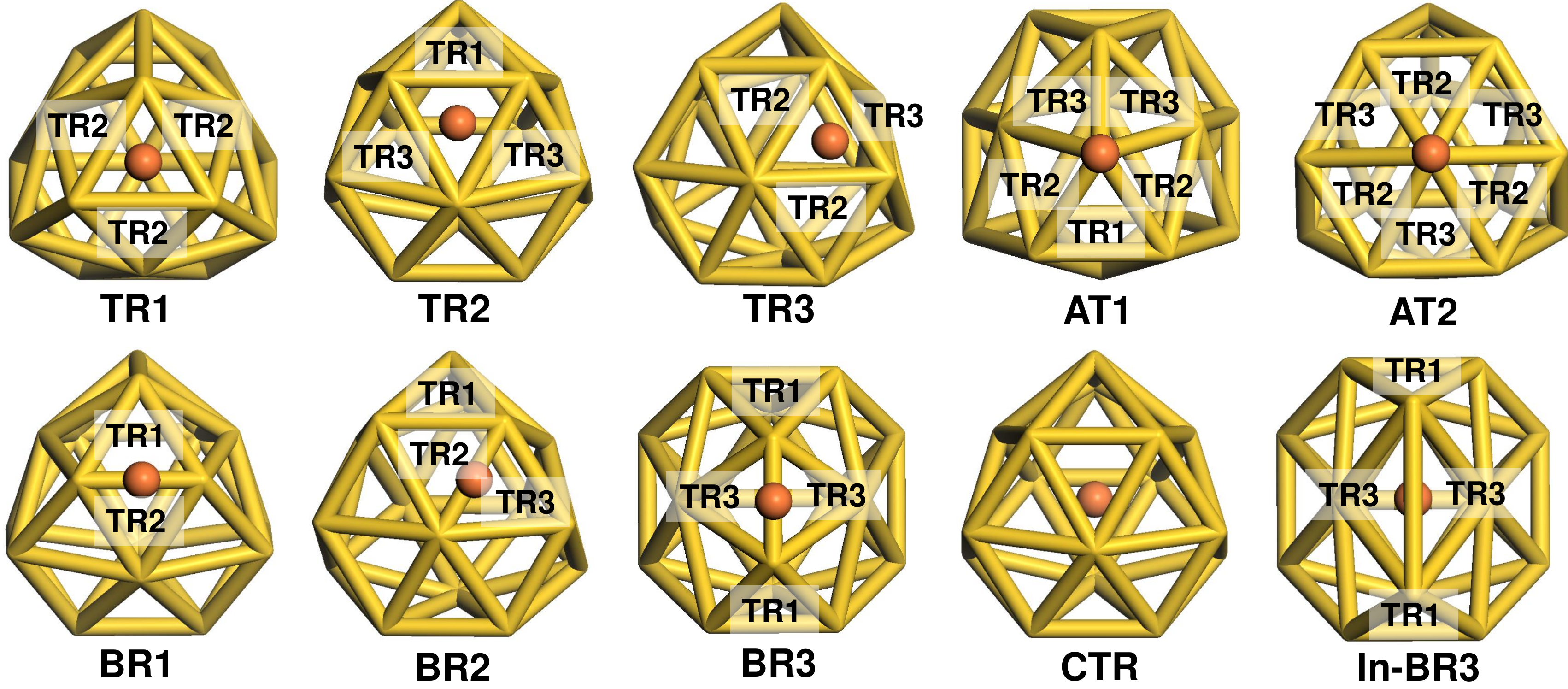}
\caption{Atomic model structures of the nitrogen atom adsorbed on
Au$_{16}^{-}$. Ten adsorption sites of the adsorbate are considered:
three different triangular sites, TR1, TR2, and TR3; three dissimilar
bridge sites, BR1, BR2, and BR3; two inequivalent atop sites, AT1 and
AT2; and two endohedral doping sites, CTR and In-BR3. For the detailed
classification of these adsorption sites, see the text.}
\label{Fig1}
\end{figure*}

To study the binding properties of a nitrogen atom on Au$_{16}^-$, we
performed first-principles calculations within DFT with spin
polarization implemented in the DMol$^3$ package~\cite{JCP90}. We used
the generalized gradient approximation with the Perdew-Burke-Ernzerhof
functional~\cite{PBE} to describe the exchange-correlation functional.
To expand the electronic wavefunctions, a double-numerical polarized
basis set was chosen with a real space cutoff of 6.5~\AA. The scalar
relativistic effects were included in the all-electron calculation. For
more precise computations, we chose an octupole scheme for the
multipolar fitting procedure on the charge density and a fine grid
scheme for numerical integration~\cite{Baerends73}.

The cluster geometry was optimized by the
Broyden-Fletcher-Goldfarb-Shanno algorithm~\cite{BCG,FR1,GD,SDF,SDFJP}
without symmetry constraints until the total energy change was
converged to $10^{-5}$~Ha in the self-consistent loop and the atomic
forces were smaller than $4\times10^{-4}$~Ha \AA$^{-1}$. The optimized
geometries of NAu$_{16}^{-}$ were determined from the total energy
minimum, and were further verified by the absence of imaginary
frequencies in the harmonic frequency calculations. Using a nudged
elastic band (NEB) method~\cite{NEB}, we searched for the
minimum-energy path in geometrical phase space connecting two stable
structures. Including the initial and final configurations, twenty
replicas were chosen to construct the NEB. For a trajectory of the
minimum-energy path between reactants and products during the atomic
encapsulation, linear synchronous transit and quadratic synchronous
transit (LST/QST) calculations~\cite{JCC86,JCP81} were also performed
with conjugate gradient minimization~\cite{congra}. Finally, IR-active
vibrational spectra were evaluated for our models by calculating the
dynamical matrix~\cite{MV80} and the Born effective charge
tensor~\cite{Baroni}.

The golden cage Au$_{16}^-$ is a truncated tetrahedron with
tetrahedral symmetry. As in our previous
study~\cite{SHKang11}, we considered various endohedral and exohedral
adsorption sites for the initial configurations, as shown in
Figure~\ref{Fig1}. A truncated tetrahedron is composed of four
truncated triangular facets (defined as TR1)~\cite{SHKang11} and four
hexagonal facets. In particular, each hexagonal facet in Au$_{16}^-$
consists of six triangles as shown in Figure~\ref{Fig1}. They are
geometrically different from TR1, and are classified into two types,
TR2 and TR3. TR2 shares an edge with an adjacent TR1 and two edges with
two TR3 triangles in the same hexagonal facet, whereas TR3 has a common
junction with another TR3 located on a neighboring hexagonal facet, and
two junctions with two TR2 triangles in the same hexagonal facet. Three
TR2 and three TR3 triangles are arranged alternately on each hexagonal
facet. Edges shared by two neighboring triangles or bridges between two
Au atoms can be classified into three types, BR1, BR2, and BR3. BR1 is
shared by TR1 and TR2; BR2 is done by TR2 and TR3 in the same hexagonal
facet; and BR3 by two TR3 triangles, one in a hexagonal facet and the
other in a neighboring hexagonal facet. Besides, two types of vertices
are considered (atop sites): AT1 and AT2. AT1 represents twelve
equivalent vertices forming four TR1 triangles, and the remaining four
vertices located at the center of each hexagonal facet fall into the
other type, AT2. Overall, there exist eight different adsorption sites
on the outer surface of Au$_{16}^-$ as shown in
Figure~\ref{Fig1}.

We also considered the encapsulation inside Au$_{16}^-$ as
``adsorption'' sites for endohedral doping of the N
atom~\cite{SHKang11}. Together with the center (CTR) of the
Au$_{16}^{-}$ cluster, eight more off-centered encapsulation
configurations named as ``In-XX{\em{n}}'' corresponding
to the eight exohedral sites (XX{\em{n}}, XX=TR, BR, or AT). Our DFT
calculations revealed that only CTR and In-BR3 sites are stable among
all endohedral doping sites, and the other seven adsorption
configurations are unstable and tend to change into either In-BR3 or
CTR during structural relaxation.

\section{Results and discussion}
\label{Results}

\begin{table}[b]
\centering
\caption{The binding energy ($E_{\rm{b}}$), the HOMO-LUMO gap
($E_{\rm{gap}}$), the distance between the nitrogen atom and its
nearest gold atoms in the golden cage ($d_{\rm{Au-N}}$), and the amount
of electronic charge transferred from Au$_{16}^-$ to the adsorbate atom
(${\Delta}Q$) at the different adsorption sites shown in Figure~\ref{Fig1}.
\label{table1}}
\begin{tabular}{ccccccc}
\hline
        & \multicolumn{3}{c} {N on Au$_{16}^-$} \\
\hline
 configuration & $E_{\rm b}$ (eV) & $E_{\rm {gap}}$ (eV) & $d_{\rm Au-N}$ (\AA) & ${\Delta}Q$ ($e$) \\
\hline
 TR1    & 2.95 & 0.14  & 1.98 & $-$0.513  \\
 TR2    & 2.80 & 0.83  & 2.03 & $-$0.477  \\
 TR3    & 1.73 & 0.13  & 2.36 & $-$0.486  \\
 BR2    & 2.57 & 0.32  & 1.94 & $-$0.404  \\
 BR3    & 2.90 & 0.64  & 1.91 & $-$0.439  \\
 AT1    & 1.68 & 0.57  & 1.84 & $-$0.353  \\
 AT2    & 1.12 & 0.68  & 1.87 & $-$0.323  \\
 CTR    & 0.58 & 0.69  & 2.20 &  $-$0.142  \\
 In-BR3 & 1.95 & 0.26  & 2.05 & $-$0.630 \\
\hline
\end{tabular}
\end{table}

For each configuration shown in Figure~\ref{Fig1}, we obtained the
equilibrium geometry of NAu$_{16}^-$ and its binding energy
($E_{\rm b}$) defined by
\[ 
    E_{\rm b}=E[\mbox{Au}_{16}^-]+E[\mbox{N}]-E[\mbox{NAu}_{16}^-],
\]
where $E[\mbox{NAu}_{16}^-]$ and $E[\mbox{Au}_{16}^-]$ are the total
energies of the golden cage with and without an N atom, respectively,
and $E$[N] represents the energy of an isolated N atom. All the
calculated energy values are summarized in Table~\ref{table1}. Here, a
positive (negative) sign in $E_{\rm b}$ means an exothermic
(endothermic) process. We found that two triangular sites (TR1 and TR2)
and one bridge site (BR3) are favored by the N atom. TR1 is the most
preferred adsorption site with a binding energy of 2.95 eV, and BR3
(TR2) has a binding energy of 2.90 eV (2.80~eV). The binding energy for
the TR3 site is not as large as those for TR1 and TR2, as listed in
Table~\ref{table1}. By contrast, the atop sites (AT1 and AT2) are less
preferable ($\approx$1.7~eV at AT1 and $\approx$1.1~eV at AT2). The N
atom that was initially located at BR1 spontaneously moved to TR2
during structural relaxation, implying that BR1 is an unstable
adsorption site.

We also calculated the encapsulation energy, which is regarded as an
energy gain due to endohedral doping. For the CTR site, the
encapsulation energy for the nitrogen atom is 0.58~eV, which is much
smaller than the binding energies on all the exohedral adsorption
sites. Intriguingly, the only stable off-centered adsorption site,
In-BR3, shows much bigger encapsulation energy (1.95~eV) than CTR. As
shown in Figure~\ref{Fig1}, in the In-BR3 case, the endoherally-doped N
atom appears to elongate the Au-Au bond corresponding to BR3 and tend
to pull in two Au atoms corresponding to AT2, compared to its exohedral
counterpart, BR3.

\begin{figure}[t]
\centering
\includegraphics[width=0.9\columnwidth]{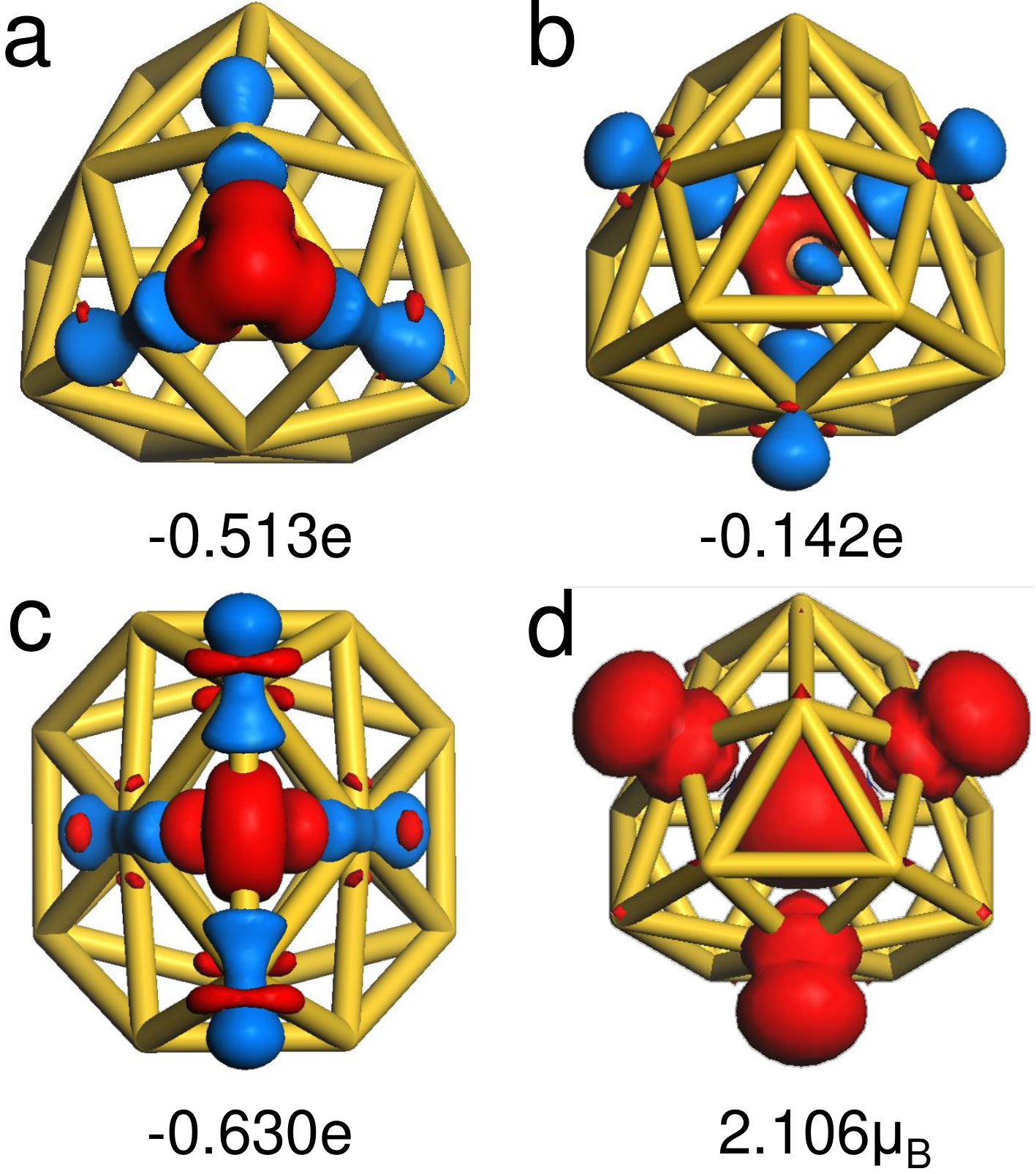}
\caption{The isosurface plots of charge density difference for (a) the
TR1, (b) the CTR, and (c) the In-BR3 structures. The electron
accumulation is represented in red, and the depletion in blue. The
negative value shown at the bottom of each structure represents the
amount of donated electrons from the golden cage to the N atom, which
evaluated using the Mulliken population analysis. (d) The spin density
plot of the CTR structure. Its net magnetic moment is calculated to be
2.106~$\mu_B$, where $\mu_B$ is the Bohr magneton. The isovalue
is ${\pm}0.005$~$e$~\AA$^{-3}$.}
\label{Fig2}
\end{figure}

Using the charge density difference, we investigated the charge
redistribution induced by the N atom adsorption.
Figures~\ref{Fig2}a--c displays the charge density differences of
TR1, the most stable exohedral adsorption site, and CTR and In-BR3, two
stable endohedral adsorption sites. The charge differences reveal
dumbbell-like shapes with two lobes at the positions of adjacent gold
atoms. The charge transfer, ${\Delta}Q$ between the adsorbate and
the golden fullerene was also calculated using the Mulliken population
analysis. As listed in Table~\ref{table1}, electrons are donated from
Au$_{16}^-$ to the N atom for all the configurations.
CTR shows the smallest amount of charge transfer, whereas In-BR3 does
the largest charge transfer. We also analyzed the local magnetic
moments of our model systems. It was found that among the adsorption
configurations we considered, only CTR has the spin magnetic moment of
$\approx$2.1~$\mu_B$, where $\mu_B$ is the Bohr magneton
($\approx5.788\times10^{-5}$~eV T$^{-1}$, where T denotes the tesla,
the unit of magnetic flux density). Figure~\ref{Fig2}d shows the spin
density plot of CTR, where the majority spin density is distributed
only at Au atoms corresponding to AT2 as well as at the N atom at the
center. The minority spin density does not appear with the isovalue of
${\pm}0.005$~$e$~\AA$^{-3}$.

\begin{figure}[t]
\centerline{\includegraphics[width=0.9\columnwidth]{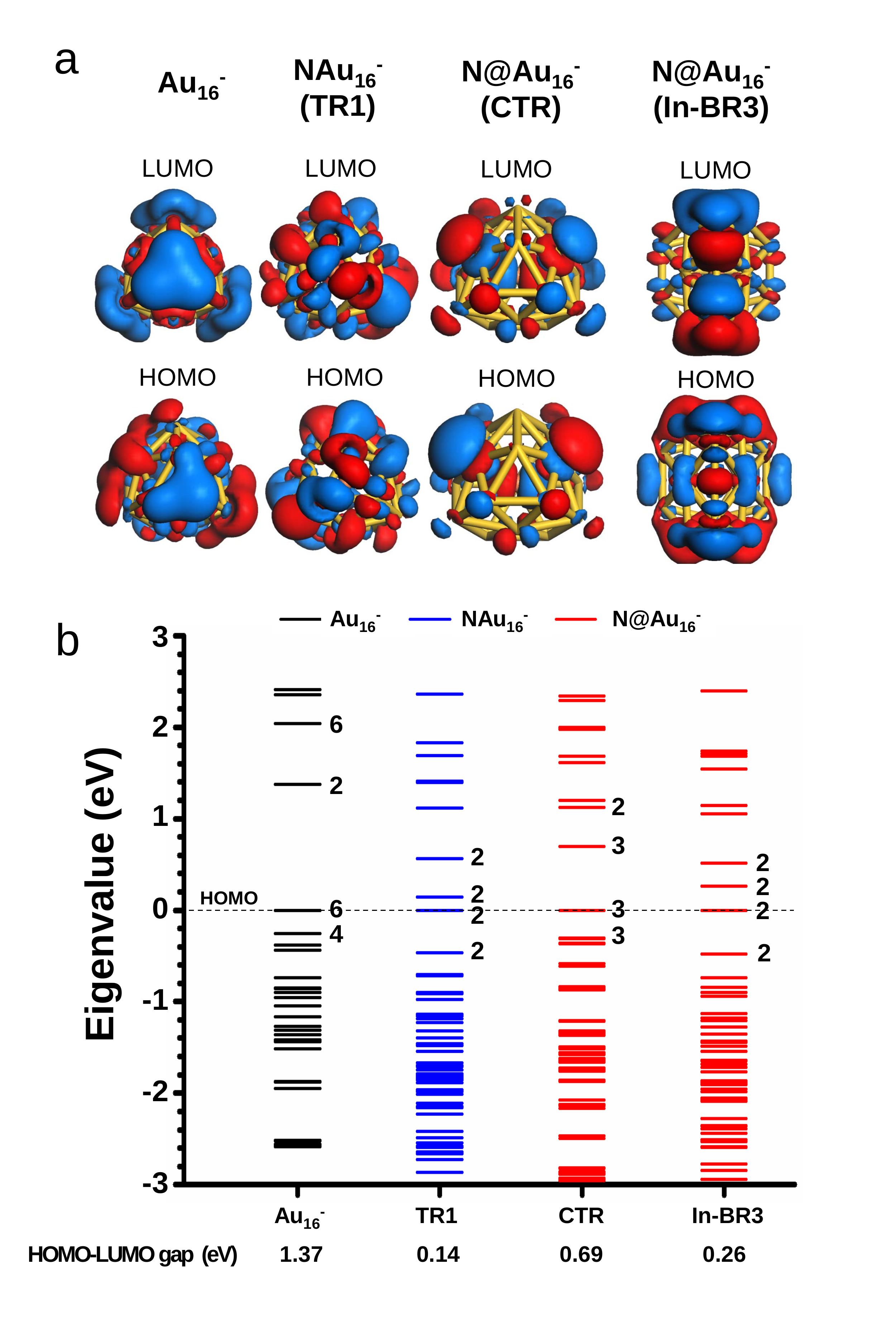}}
\caption{(a) HOMO and LUMO of the bare Au$_{16}^-$ cage, the most
stable exohedrally adsorbed structure (TR1) of NAu$_{16}^-$, and the
N-encapsulated structures (CTR and In-BR3) of N@Au$_{16}^-$. The
isovalue is ${\pm}0.02$~$e$~\AA$^{-3}$. (b) Energy levels near the
HOMOs and the LUMOs of the four structures given in (a). The numbers
written next to the energy levels represent the degeneracy of
respective levels. The HOMO level of every structure is set to zero.
For each configuration, the HOMO--LUMO energy gap is given at the
bottom of the graph.}
\label{Fig3}
\end{figure}

Next, we move on to the energy eigenvalues and the corresponding
molecular orbitals (eigenfunctions). Figure~\ref{Fig3}a displays the
highest occupied molecular orbitals (HOMOs) and lowest unoccupied
molecular orbitals (LUMOs) of the three configurations shown in
Figures~\ref{Fig2}a--c as well as those of the bare golden cage for
comparison. Obviously, we can find the antibonding character of N 2$p$
orbitals with Au 5$d$ orbitals. The adsorption of the nitrogen atom
causes the breaking of the tetrahedral symmetry, and some degeneracy is
lifted resulting in level splitting as shown in Figure~\ref{Fig3}b. The
HOMO--LUMO energy gaps are calculated to be 0.14~eV, 0.69~eV, and
0.26~eV for TR1, CTR, and In-BR3, respectively, all of which are much
smaller than that (1.37~eV) of the bare Au$_{16}^-$. Because of its much
smaller HOMO--LUMO gap, the exohedrally doped complex TR1 may be
chemically more reactive than the endohedrally doped configurations.

\begin{figure}[t]
\centering
\includegraphics[width=0.9\columnwidth]{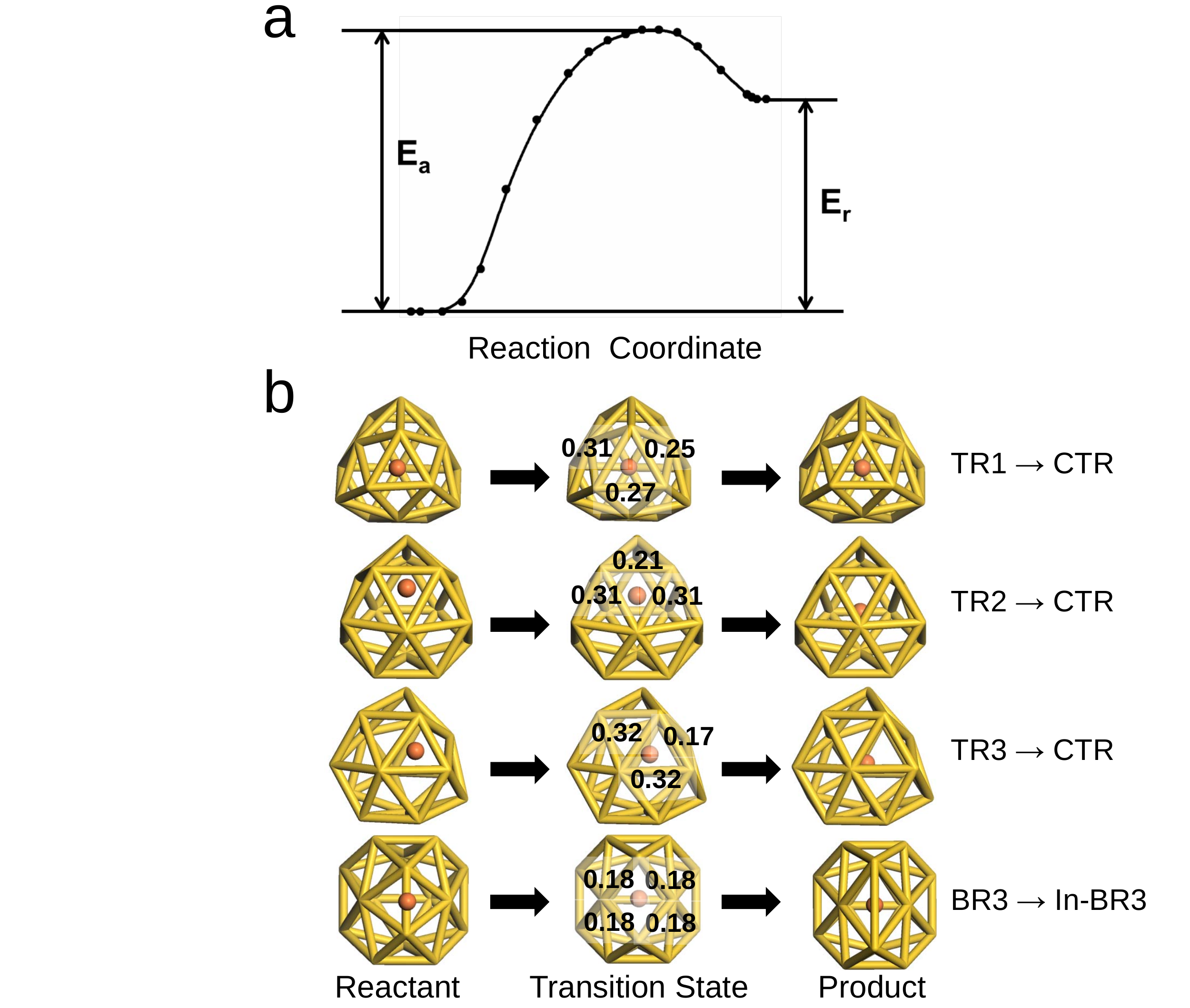}
\caption{(a) Schematic energy profile along the reaction coordinates.
The dots represent the energies of the replicas in the transformation
in the NEB method. The activation energy barrier $E_{\rm{a}}$ and the
reaction energy $E_{\rm{r}}$ are defined by the encapsulation barrier
from the reactant and the energy difference between the reactant and
the product, respectively. (b) Atomistic model structures for various
reaction paths from reactants to products. The first three figures
display the transitions of the N atom from three triangular sites (TR1,
TR2, and TR3) to the CTR site, whereas the last one depicts that from
BR3 to In-BR3, the off-centered site.The values marked in the middle
represent the maximal elongations ${\Delta}d_{\rm{Au-Au}}$ in \AA from
the equilibrium distances between two neighboring Au atoms located near
the N atom during the encapsulation process.
\label{Fig4}}
\end{figure}

\begin{table}[b]
\centering
\caption{The activation energy barriers $E_{\rm{a}}$ and the reaction
energies $E_{\rm{r}}$ for the four transitions described in
Figure~\ref{Fig4}.
\label{table2}}
\begin{tabular}{ccccc}
\hline
\multicolumn {2}{c} {System} & $E_{\rm{a}}$ (eV) & $E_{\rm{r}}$ (eV) \\
\hline
             & TR1 $\rightarrow$ CTR & 2.71 & 2.37   \\
NAu$_{16}^-$ & TR2 $\rightarrow$ CTR & 2.71 & 2.22   \\
             & TR3 $\rightarrow$ CTR & 2.74 & 1.15   \\
             & BR3 $\rightarrow$ In-BR3 & 1.51 & 0.94 \\
\hline
\end{tabular}
\end{table}

To evaluate the activation energy barrier associated with nitrogen atom
encapsulation, which plays a pivotal role in the reaction rate of the
dopant insertion, we obtained the minimum-energy path in the
configuration space using the NEB method. Figure~\ref{Fig4}a is
an illustration of the energy profile in the transformation between the
stable exohedral and endohedral adsorption configurations, showing the
activation barrier ($E_{\rm{a}}$), and reaction energy ($E_{\rm{r}}$)
defined by the energy difference between the reactant and the product.
Because any triangular site can be a gate for atomic insertion, we
first considered three paths: TR1 $\rightarrow$ CTR, TR2 $\rightarrow$
CTR, and TR3 $\rightarrow$ CTR. In these cases, three Au-Au bonds are
maximally stretched in the transition states during insertion of the
dopant atom. We found that the activation energy barriers of the N atom
along all the three paths are almost the same ($\approx2.7$~eV). These
similar activation barriers are strongly related to the maximally
elongated Au-Au bond lengths, which are essentially the same in all
three cases, as displayed in the middle column of Figure~\ref{Fig4}b
representing the transition states. On the other hand, the reaction
energy of the TR3 $\rightarrow$ CTR path is about 1.2~eV smaller than
the other two paths, since TR3 has a smaller binding energy with a
longer Au-N distance than TR1 and TR2 (see Table \ref{table1}). We also
checked the BR3 $\rightarrow$ In-BR3 path, since BR3 and In-BR3 are
also stable exohedral and endohedral configurations. Interestingly, for
the BR3 $\rightarrow$ In-BR3 path, a single Au-Au bond corresponding to
BR3 is broken to form Au-N-Au and at the same time the N atom tend to
draw two Au atoms located at two nearest AT2 sites. In this transition,
the activation energy barrier is much smaller to be 1.51 eV than those
of three paths through the triangular sites. In addition, much more
stable In-BR3  configuration than CTR for the N encapsulation shows
much lower reaction energy, which is about 0.94~eV. Consequently, we
conclude that the transition path occurring through a single Au-Au bond
stretch may be preferred in the N atom encapsulation process. The
calculated activation barriers ($E_{\rm{a}}$) and reaction energies
($E_{\rm{r}}$) are summarized in Table~\ref{table2}.

The encapsulation events through any triangular site would not take
place even near the melting temperature of the golden fullerene, while the
insertion event though the BR3 site may occur at relatively low
temperatures without any external aids. According to our estimation,
the probabilities that such transitions happen are about
$\sim10^{-11}$~s$^{-1}$ for TR1 $\rightarrow$ CTR, and
$\sim10^{-3}$~s$^{-1}$ for BR3 $\rightarrow$ In-BR3 at 600~K. The
insertion events would be even enhanced with the help of light or
electron-beam irradiation as shown in earlier experiments. It was shown
that electron-beam irradiation has been used to fuse C$_{60}$
fullerenes inside host carbon nanotubes~\cite{Bando01} and the scanning
tunneling microscope has induced a cis-trans conformation change in the
azobenzene molecule~\cite{BChoi06}.

\begin{figure}[t]
\centering
\includegraphics[width=0.9\columnwidth]{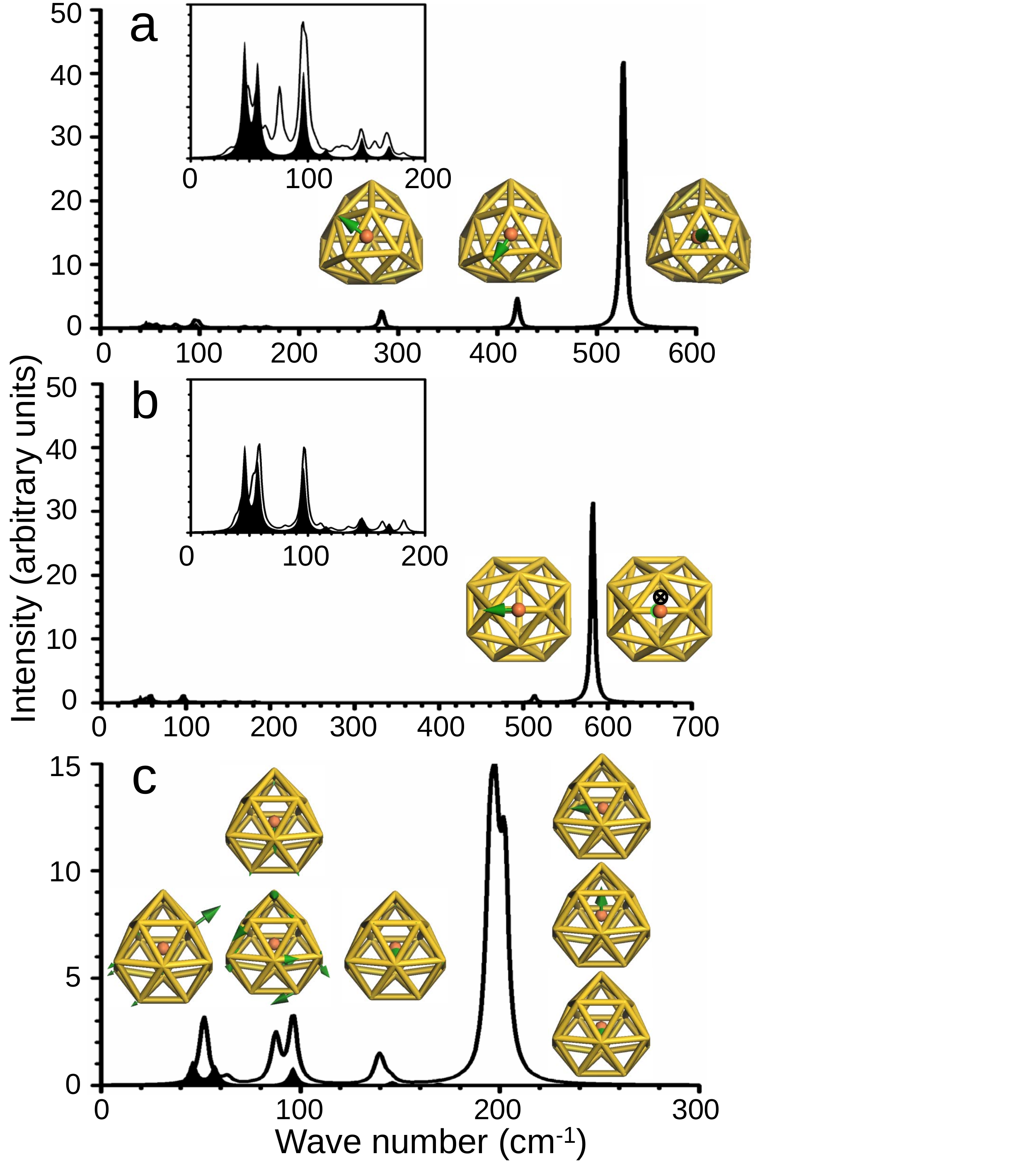}
\caption{The IR-active modes for (a) the most stable exohedral
adsorption structure (TR1), (b) the next most stable exohedral
adsorption structure (BR3), and (c) an endohedrally doped N@Au$_{16}^-$
(CRT). The IR-active modes for the bare Au$_{16}^-$ cage are displayed
with solid black in each figure. Detailed IR spectra at wavenumbers below
200~cm$^{-1}$ are shown in the insets for the exohedral cases. The
model structures with green arrows are shown at the peaks to denote the
directions and the intensities of the IR-active modes contained in each peak.
In (b), the symbol $\otimes$ indicates that the nitrogen atom moves
towards the center of the Au-Au bond.
\label{Fig5}}
\end{figure}

Finally, we calculated the IR active vibrational spectra of two most
stable exohedral doping configurations (TR1 and BR3), and an endohedral
configuration (CRT). We believe that different IR spectra could be used
to distinguish structures with different adsorption sites.
Figure~\ref{Fig5} shows our computationally simulated IR intensity data
for the three configurations. For comparison, we also calculated the IR
spectrum for the bare Au$_{16}^-$ cage, displayed in solid black in
each figure. Its strong IR intensities are observed in the frequency
range below 200~cm$^{-1}$, since Au is one of heavy elements, whose
frequency is inversely proportional to $\sqrt{m}$, where $m$ is the
atomic mass. We found that three modes with dominant intensities arise
at frequencies of 45, 56, and 96~cm$^{-1}$ corresponding to an Au-Au
bond stretching mode and several weak modes. In the case of the TR1
configuration, the strongest peak appears at much higher frequency of
$\approx520$~cm$^{-1}$ as well as two weak peaks at
$\approx280$~cm$^{-1}$ and $\approx420$~cm$^{-1}$ due to the N
adsorbate, as shown in Figure~\ref{Fig5}a. For the BR3 configuration,
the strongest peak arises at even higher frequency of
$\approx580$~cm$^{-1}$ with a very weak peak near
$\approx520$~cm$^{-1}$. It means that BR3 is spectroscopically
distinguishable from TR1. The direction and the relative intensity of each
peak are denoted by the green arrows.

On the other hand, the insertion of the N atom at the
center combines two lowest modes (45 and 56~cm$^{-1}$) of the bare
golden cage into one mode while keeping the mode around 96~cm$^{-1}$,
as displayed in Figure~\ref{Fig5}c. Three other peaks are observed near
90, 140 and 200~cm$^{-1}$. As denoted by the green arrows in
Figure~\ref{Fig5}c, all of them are due to the vibration motion of the
N atom encapsulated at the center of the golden fullerene. Especially, the
peak at 200~cm$^{-1}$ is composed of three IR-active modes. These
highest IR-active modes occur at much lower frequency than those of
either TR1 or BR3. This is attributed to the flat potential surface of
the N atom near the center of Au$_{16}^-$ resulting in a small
force constant.

\section{Conclusion}
\label{Conclusion}

In summary, we studied the binding effects of a nitrogen atom on the
structural and electronic properties of an anionic golden fullerene
Au$_{16}^-$ using density functional theory. The general trend is that
the triangular (hollow) and bridge sites are preferred for exohedral
nitrogen adsorption. When the N atom is adsorbed on the cage, electrons
are transferred from Au$_{16}^-$ to the N atom, and the binding
character tends to be ionic. We also searched for the minimum-energy
transition paths between the stable exohedral and endohedral adsorption
configurations, and obtained the activation energy and reaction energy
for each path. It was found that the transitions from the triangular
sites to the center of Au$_{16}^-$ would not take place with an
activation energy of about 2.7~eV and a reaction energy of $1.2-2.4$~eV
implying no such transition without an external aid. On the other hand,
we found that the transition through a Au-Au bond may take place even
at ambient temperatures, since the transition energy barrier is much
lower. The IR spectra for various adsorption configurations were also
computed. In the exohedral doping cases, the dominant IR peaks occur at
higher frequencies with larger intensities than in the endohedral
doping cases. Our study may help design new types of small gold
nanocluster derivatives. On the basis of our present work, further
studies will be carried out on the binding properties of
nitrogen-containing molecules such as N$_2$, NO, NO$_2$ and NH$_3$ on
the golden cage.

\section*{Acknowledgement}
G.K. thanks the Basic Research program (Grant No. 2012-0001743) and the
Priority Research Center Program (Grant No. 2012-0005859) of the Korean
Government Ministry of Education, Science and Technology. Y.K.
gratefully acknowledges the financial support from the National
Research Foundation of Korea (Grant Nos. 2011-0002456 and
2011-0016188). Some portion of our computational work was done using
the resources of the KISTI Supercomputing Center (KSC-2011-C1-04 and
KSC-2011-C1-19).


\begin{thebibliography}{10}
\expandafter\ifx\csname url\endcsname\relax
  \def\url#1{{\tt #1}}\fi
\expandafter\ifx\csname urlprefix\endcsname\relax\def\urlprefix{URL }\fi
\providecommand{\eprint}[2][]{\url{#2}}

\bibitem{HIVFD}
S.H. Friedman, D.L. DeCamp, R.P. Sijbesma, G. Srdanov, F. Wudl,
G.L. Kenyon, J. Am. Chem. Soc. 115 (1993) 6506.

\bibitem{FFFD}
M. Prato, M. Maggini,
Acc. Chem. Res. 31 (1998) 519.

\bibitem{EDFD}
L. Echegoyen, L.E. Echegoyen,
Acc. Chem. Res. 31 (1998) 593.

\bibitem{ESFD}
D.M. Guldi, M. Prato,
Acc. Chem. Res. 33 (2000) 695.

\bibitem{Bulusu06}
S. Bulusu, X. Li, L.S. Wang, X.C. Zheng,
Proc. Natl. Acad. Sci. U.S.A. 103 (2006) 8326.

\bibitem{Kohn}
P. Hohenberg, W. Kohn,
Phys. Rev. 136 (1964) B864.

\bibitem{PCCP06}
M. Walter, H. Hakkinen,
Phys. Chem. Chem. Phys 8 (2006) 5407.

\bibitem{JCP07Si}
Q. Sun, Q. Wang, G. Chen, P. Jena,
J. Chem. Phys. 127 (2007) 214706.

\bibitem{ANGE07CU}
L.M. Wang, S. Bulusu, H.J. Zhai, X.C. Zeng, L.S. Wang,
Angew. Chem., Int. Ed. 46 (2007) 2915.

\bibitem{PLA08}
W. Fa, A. Yang,
Phys. Lett. A 372 (2008) 6392.

\bibitem{Yoon03}
B. Yoon, H. H\"kkinen, U. Landman,
J. Phys. Chem. A 107 (2003) 4066.

\bibitem{Valden98}
M. Valden, X. Lai, D.W. Goodman,
Science 281 (1998) 1647.

\bibitem{Heiz00}
U. Heiz, W.D. Schneider,
J. Phys. D: Appl. Phys. 33 (2000) R85.

\bibitem{Boyen02}
H.-G. Boyen, G. K\"astle, F. Weigl, B. Koslowski, C. Dietrich, P. Ziemann, J. P. Spatz, S. Riethm\"uller, C. Hartmann, M. M\"oller, G. Schmid,
M. G. Garnier and P. Oelhafen,
Science 297 (2002) 1533.

\bibitem{SHKang11}
S.-H. Kang, G. Kim, Y.-K. Kwon,
J. Phys.: Condens. Matter 23 (2011) 505301.

\bibitem{JCP90}
B. Delley,
J. Chem. Phys. 92 (1990) 508.

\bibitem{PBE}
J.P. Perdew, K. Burke, M. Ernzerhof,
Phys. Rev. Lett. 77 (1996) 3865.


\bibitem{Baerends73}
E.J. Baerends, D.E. Ellis, P. Ros,
Chem. Phys. 2 (1973) 41.

\bibitem{BCG}
C. G. Broyden,
J. Inst. Math. Appl. 6 (1970) 76.

\bibitem{FR1}
R. Fletcher,
Comput. J. 13 (1970) 317.

\bibitem{GD}
D. Goldfarb,
Math. Comput. 24 (1970) 23.

\bibitem{SDF}
D.F. Shanno,
Math. Comput. 24 (1970) 647.

\bibitem{SDFJP}
D.F. Shanno, P.C. Kettler,
Math. Comput. 24 (1970) 657.

\bibitem{NEB}
G. Henkelman, H. Jonsson,
J. Chem. Phys. 113 (2000) 9978.

\bibitem{JCC86}
J. Baker,
J. Comput. Chem. 7 (1986) 385.

\bibitem{JCP81}
C.J. Cerjan, W.H. Miller,
J. Chem. Phys. 75 (1981) 2800.

\bibitem{congra}
N. Govind, M. Petersen, G. Fitzgerald, D. King-Smith, J. Andzelm,
Comput. Mater. Sci. 28 (2003) 250.

\bibitem{MV80}
E.B. Wilsun, J.C. Decius, P.C. Cross,
Molecular Vibrations (New York: Dover, 1980).

\bibitem{Baroni}
S. Baroni, S. de Gironcoli, A. dal Corso, P. Giannozzi,
Rev. Mod. Phys. 73 (2001) 515.

\bibitem{Bando01}
S. Bando, M. Takizawa, K. Hirahara, M. Yudasaka, S. Iijima,
Chem. Phys. Lett. 337 (2001) 48.

\bibitem{BChoi06}
B. Choi, S. Kahng, S. Kim, H. Kim, H.W. Kim, Y.J. Song, J. Ihm, Y. Kuk,
Phys. Rev. Lett. 96 (2006) 156106.

\end{thebibliography}
\providecommand{\newblock}{}

\end{document}